\documentclass[english]{article}
\usepackage[LGR,T1]{fontenc}
\usepackage[latin9]{inputenc}
\usepackage{geometry}
\usepackage{float}
\usepackage{textcomp}
\usepackage{amsmath}
\usepackage{amssymb}

\makeatletter

\DeclareRobustCommand{\greektext}{%
  \fontencoding{LGR}\selectfont\def\encodingdefault{LGR}}
\DeclareRobustCommand{\textgreek}[1]{\leavevmode{\greektext #1}}
\ProvideTextCommand{\~}{LGR}[1]{\char126#1}

\makeatother

\usepackage{babel}
\begin{document}
\title{Eigenvalue crossings in equivariant families of matrices}
\author{Jonathan Rawlinson}
\maketitle

\begin{abstract}
    According to a result of Wigner and von Neumann \cite{wignervonneumann}, real symmetric matrices with a doubly degenerate lowest eigenvalue form a submanifold of codimension $2$ within the space of all real symmetric matrices. This mathematical result has important consequences for chemistry. First, it implies that degeneracies do not occur within generic one-parameter families of real symmetric matrices - this is the famous non-crossing rule, and is responsible for the phenomenon of avoided crossings in the energy levels of diatomic molecules. Second, it implies that energy levels are expected to cross in polyatomic molecules, with crossings taking place on a submanifold of nuclear configuration space which is codimension $2$ - this submanifold is the famous conical intersection seam, of central importance in nonadiabatic chemistry.
    In this paper we extend the analysis of Wigner and von Neumann to include symmetry. We introduce a symmetry group, and consider parametrised families of matrices which respect an action of that symmetry group on both the parameter space and on the space of matrices. A concrete application is to molecules, for which the relevant symmetry group is generated by permutations of atomic nuclei combined with spatial reflections and rotations. In the presence of this extra symmetry, we find that energy level crossings do not typically occur on codimension $2$ submanifolds, and connect our findings with the discovery of confluences of conical intersection seams in the chemical literature. We give a classification of confluences for triatomic molecules and planar molecules, unifying the previous literature on this topic, and predict several new types of confluence. 
    
\end{abstract}

\section{Introduction}

This paper is concerned with real symmetric matrices depending on
parameters. These can be described by a map
\begin{equation}
f:P\to M\label{eq:fund_map}
\end{equation}
from a parameter space $P$ to the space of real symmetric $n\times n$
matrices $M$. For each $x\in P$, we have a matrix $f\left(x\right)\in M$
with a corresponding set of real eigenvalues
\[
\lambda_{1}\left(x\right)\leq\lambda_{2}\left(x\right)\leq\ldots\leq\lambda_{n}\left(x\right).
\]
Suppose $x^{*}\in P$ is such that 
\[
\lambda_{1}\left(x^{*}\right)=\lambda_{2}\left(x^{*}\right)\neq\lambda_{3}\left(x^{*}\right).
\]
That is, $f\left(x^{*}\right)$ has a doubly degenerate lowest eigenvalue.
Starting at $x^{*}$, as $x$ varies, $f\left(x\right)$ varies and
so does its eigenvalues. Most variations break the double degeneracy.
However, the double degeneracy persists on some subset $\Sigma\subset P$
containing $x^{*}$. What does $\Sigma$ look like near $x^{*}$?

Early work on this problem was motivated by quantum mechanics and
its application to molecules. In Born-Oppenheimer theory, one is interested
in the energy eigenvalues of the electronic Hamiltonian, $\hat{H}_{e}\left(x\right)$,
a self-adjoint operator on the electronic Hilbert space. Each $x\in P$
corresponds to a particular fixed configuration of the atomic nuclei,
and the Hamiltonian $\hat{H}_{e}\left(x\right)$ describes (amongst
other things) the electrostatic attraction of the electrons towards
those nuclei. As $x$ varies, so do the eigenvalues of the electronic
Hamiltonian, tracing out potential energy surfaces:
\[
E_{0}\left(x\right)\leq E_{1}\left(x\right)\leq\ldots
\]
where $E$ stands for energy. Often one is interested only in $E_{0}\left(x\right)$,
the ground state potential energy surface. This is because many chemical
phenomena can be described, to a good approximation, in terms of motion
of atomic nuclei governed by the single surface $E_{0}\left(x\right)$.
Such an approximation breaks down, however, in regions of nuclear
configuration space surrounding eigenvalue degeneracies. 

Wigner and von Neumann showed in $1929$ that \emph{two} parameters
must be varied in order to arrange an eigenvalue degeneracy in a real
symmetric matrix. Another way of saying this is that if there is a
double degeneracy at $x^{*}\in P$, then locally $\Sigma\subset P$
(the subset on which the double degeneracy persists) is a submanifold
of codimension $2$. The immediate application of Wigner and von Neumann's
result was to diatomics. In the case of only two atoms, the shape
of the molecule is described by just one parameter (the interatomic
distance). Consequently, $\Sigma$ (being codimension $2$) is empty
so electronic energy levels never cross in diatomic molecules \cite{wignervonneumann,landau_lifschitz}.
On the other hand, in polyatomic molecules, the molecular shape is
described by three or more degrees of freedom so eigenvalue degeneracies
are indeed expected to occur on non-trivial submanifolds $\Sigma\subset P$
of codimension $2$. These submanifolds have been studied extensively
in the chemical literature and are known as \emph{seams of conical
intersection}.

The situation changes if the physical system of interest has some
additional \emph{symmetry}. Given an $n$-dimensional representation
of a compact Lie group $G$ by matrices $\Gamma\left(g\right)$ (each
$g\in G$), one considers maps of the form (\ref{eq:fund_map}) which
additionally satisfy
\[
f\left(x\right)=\Gamma\left(g\right)f\left(x\right)\Gamma\left(g\right)^{-1},\hspace{1em}\forall g\in G,\hspace{1em}\forall x\in P.
\]
This last equation says that the action of $G$ commutes with $f\left(x\right)$.
In particular, the eigenspaces of $f\left(x\right)$ form representations
of $G$. Suppose $x^{*}\in P$ is such that the lowest eigenspace
is a sum of \emph{two} irreducible representations (the analogue of
double degeneracy):
\[
\ker\left(f\left(x^{*}\right)-\lambda_{1}\left(x^{*}\right)\right)\cong R_{1}\oplus R_{2}
\]
where $R_{1}$ and $R_{2}$ are representations of $G$ which are
irreducible (over the real numbers). Starting at $x^{*}$, most variations
of $x$ cause a splitting into $R_{1}$ and $R_{2}$ eigenspaces having
distinct eigenvalues. However, the splitting is avoided on some subset
$\Sigma\subset P$ containing $x^{*}$. Dellnitz and Melbourne, motivated
by numerical treatment of bifurcation problems, have studied this
case in detail. They found that it is only necessary to vary one parameter
to arrange an eigenvalue degeneracy between neighbouring eigenspaces,
if those eigenspaces constitute non-isomorphic representations of
$G$. On the other hand, if the representations are isomorphic, it
is necessary to vary two, three or five parameters to arrange a degeneracy
depending on whether the relevant representation is of real, complex
or quaternionic type \cite{key-7}. Another way of saying this is
that if $R_{1}\ncong R_{2}$, then $\Sigma\subset P$ is a submanifold
of codimension $1$ near $x^{*}$; if $R_{1}\cong R_{2}\cong R$,
then $\Sigma\subset P$ is a submanifold of codimension $2$, $3$
or $5$ near $x^{*}$ depending on whether $R$ is of real, complex
or quaternionic type.

The result of Dellnitz and Melbourne generalises earlier findings
in the chemical and applied maths literature. For example, triatomic
molecules have a reflection symmetry in the plane of the molecule.
Electronic energy eigenstates are classified as $A'$-type or $A''$-type
according to whether they change sign under this reflection. It is
well-known that $A'$-$A''$ degeneracies occur on a submanifold of
codimension $1$ (since $A'$ and $A''$ are non-isomorphic), while
$A'$-$A'$ degeneracies occur on a submanifold of codimension $2$
(since $A'$ is a representation of \emph{real} type). A codimension
$3$ example is given by Arnol'd, who considers oscillations of a
membrane which has third-order rotational symmetry. Arnol'd considers
oscillations which belong to the two-dimensional non-trivial representation
of the symmetry group (which is a representation of \emph{complex}
type) and finds that it is necessary to vary \emph{three} parameters
(not just two) to arrange a degeneracy between neighbouring characteristic
frequencies \cite{arquasi}. A codimension $5$ example is given by
Mead, who shows that in Born-Oppenheimer theory \emph{five} parameters
must typically be varied in order to arrange an eigenvalue degeneracy
in systems with spin-orbit coupling and an odd number of electrons
\cite{meadnon}. The relevant group in Mead's case corresponds to time-reversal
symmetry, which is implemented on the electronic Hilbert space by
means of a map $j$ which anti-commutes with multiplication by the
complex scalar $i$ and satisfies $j^{2}=-1$. The existence of $j$
allows us to view the electronic Hilbert space (which is nominally
a complex vector space) as a \emph{real} vector space equipped with
a linear action of the quaternion group (generated by $i$ and $j$).
The Hamiltonian commutes with the action of the quaternion group (this
is the assumption of time-reversibility), so its eigenspaces are representations
of the quaternion group. In fact these eigenspaces only contain the
four-dimensional standard representation of the quaternion group (which
is a representation of \emph{quaternionic} type) so degeneracies occur
on a submanifold of codimension $5$.

The purpose of these notes is to extend the above analysis to the
case where $G$ acts not only on the space of matrices $M$, but also
\emph{on the parameter space} $P$. Given an action of $G$ on $P$
(which we write as $x\mapsto gx$), suppose we have a map of the form
(\ref{eq:fund_map}) which additionally satisfies
\[
f\left(gx\right)=\Gamma\left(g\right)f\left(x\right)\Gamma\left(g\right)^{-1},\hspace{1em}\forall g\in G,\hspace{1em}\forall x\in P.
\]
This case does not appear to have been treated in a general manner
before, yet it is a very natural extension in view of its potential
applications in molecules. In Born-Oppenheimer theory, for which the
relevant parameter space $P$ is physically the configuration space
of the atomic nuclei, physical symmetries typically act on the electronic
Hilbert space \emph{and} on the parameter space. For example, the
electronic Hamiltonian is invariant under simultaneous spatial rotations
of all nuclei and electrons. The rotation of the nuclei is an action
on $P$; the rotation of the electrons is an action on $M$. In the
chemical literature, beginning with the work of Atchity, Ruedenberg,
and Nanayakkara \cite{key-1} and then Yarkony and coworkers \cite{key-10,key-6,key-8},
it has long been appreciated that (in the presence of symmetry) $\Sigma$
is not always a submanifold, but can have a more complicated, singular
structure. Seams of conical intersection can undergo bifurcations,
giving rise to other seams, at singular points known as \emph{confluences}.
In order to understand such phenomena, it is necessary to consider
the action of $G$ on matrix space ($M$) \emph{and} parameter space
($P$).

The purpose of this paper is to give a mathematical treatment of
this extension (Sections $2$ and $3$), then apply the results to
molecules (Section $4$). In Section $2$ we analyse the case of no
symmetry (which we call the standard case). In Section $3$ we introduce
the action of a non-trivial symmetry group $G$ on parameter space
and matrix space. We call this the equivariant case. In the equivariant
case, eigenvalue degeneracies do not necessarily occur on submanifolds
in parameter space, but on subsets $\Sigma\subset P$ with more complicated
structure. We make a transversality assumption on the map $f:P\to M$,
motivated by the work of Field and Bierstone, which allows us to narrow
down the possibilities for $\Sigma$. In Section $4$ we discuss applications
to Born-Oppenheimer theory, analysing in detail the case of triatomic
molecules as well as planar configurations of more general molecules.
We explain which confluences typically arise in these cases, and connect
our work to the chemical literature. Some of the confluences we discover
are new, including an $A_{1}-A_{2}$ confluence in $D_{3h}$ symmetry
which should occur generically in triatomics formed of three identical
atoms.

\section{Standard case}

Let $\mathcal{H}$ be a real $n$-dimensional vector space with inner
product. Denote by $M$ the set of self-adjoint linear maps $\mathcal{H}\to\mathcal{H}$.
$M$ is a vector space, and we equip $M$ with an inner product given
by
\[
\langle\alpha,\beta\rangle={\rm Tr}\left(\alpha\circ\beta\right)
\]
If we choose an orthonormal basis, then elements of $M$ are represented
by real symmetric $n\times n$ matrices with respect to that basis.
Such a matrix has 
\[
N=\frac{1}{2}n\left(n+1\right)
\]
independent entries, so we can view $M$ as a Riemannian manifold
of dimension $N$.

Each map in $M$ has a corresponding set of real eigenvalues
\[
\lambda_{1}\leq\lambda_{2}\leq\ldots\leq\lambda_{n}.
\]

Consider those maps in $M$ which have a doubly degenerate lowest
eigenvalue, that is, $\lambda_{1}=\lambda_{2}\neq\lambda_{3}$. It
is a well-known result, going back to the work of Wigner and von Neumann
\cite{wignervonneumann}, that such maps form a submanifold $D\subset M$
of codimension $2$. 

The fact that $D$ is a submanifold of codimension $2$ has important
consequences in the case of maps which depend smoothly on a set of
$k$ real parameters. If we interpret these parameters as local coordinates
on a $k$-dimensional Riemannian manifold $P$, this means smooth
maps
\[
f:P\to M
\]
from some parameter manifold $P$ to the manifold of real symmetric
matrices $M$. 

Suppose $x^{*}\in P$ has $f\left(x^{*}\right)\in D$. In other words,
$f\left(x^{*}\right)$ has a double degeneracy, and we will be interested
in the set $\Sigma=f^{-1}\left(D\right)$ on which that double degeneracy
persists. Since $D$ is a submanifold of codimension $2$, we can
identify an open neighbourhood of $f\left(x^{*}\right)\in M$ with
an open neighbourhood of the origin $\left(0,0\right)$ in $\mathbb{R}^{N-2}\times\mathbb{R}^{2}$,
such that points in $D$ are identified with points in $\mathbb{R}^{N-2}\times\left\{ 0\right\} $.
We can also identify an open neighbourhood of $x^{*}\in P$ with an
open neighbourhood $\mathcal{U}$ of the origin $0\in\mathbb{R}^{k}$.
Then the local behaviour of $f$ is expressed by a map
\[
\tilde{f}:\mathcal{U}\subset\mathbb{R}^{k}\to\mathbb{R}^{N-2}\times\mathbb{R}^{2}.
\]
Define $\pi:\mathbb{R}^{N-2}\times\mathbb{R}^{2}\to\mathbb{R}^{2}$
to be projection onto the second factor, and $F=\pi\circ\tilde{f}$,
so that
\[
F:\mathcal{U}\subset\mathbb{R}^{k}\to\mathbb{R}^{2}.
\]
We have $F\left(0\right)=0$ and, locally, $\Sigma$ is given by solutions
of the equation $F=0$. $F$ is a smooth map from $\mathcal{U}\subset\mathbb{R}^{k}$
to $\mathbb{R}^{2}$. We assume that $F$ is \emph{transverse} to
$0\in\mathbb{R}^{2}$, which is a generic property of maps of this
kind. It follows that the solutions to the equation $F=0$ form a
submanifold of codimension $2$ in $\mathcal{U}\subset\mathbb{R}^{k}$. 

\section{Equivariant case}

Now suppose in addition we have a Lie group $G$ (assumed compact)
which acts on $\mathcal{H}$ by orthogonal linear maps, 
\[
\gamma\left(g\right):\mathcal{H}\to\mathcal{H}
\]
for each $g\in G$. This gives rise to a smooth action of $G$ on
$M$ according to 
\[
\alpha\mapsto\gamma\left(g\right)\circ\alpha\circ\gamma\left(g^{-1}\right)
\]
for each $\alpha\in M$ and $g\in G$. Note that $D\subset M$ is
a $G$-invariant submanifold, since eigenvalues are invariant under
conjugation.

We also introduce an action of $G$ by isometries on our parameter
manifold $P$, writing this action as
\[
x\mapsto gx
\]
for each $x\in P$ and $g\in G$. 

As before, we are interested in smooth maps
\[
f:P\to M
\]
but now we require $f$ to be $G$-equivariant:
\[
f\left(gx\right)=\gamma\left(g\right)\circ f\left(x\right)\circ\gamma\left(g^{-1}\right)
\]
for each $x\in P$ and $g\in G$. 

Suppose $x^{*}\in P$ has $f\left(x^{*}\right)\in D$. In other words,
$f\left(x^{*}\right)$ has a double degeneracy, and we will be interested
in the set $\Sigma=f^{-1}\left(D\right)$ on which that double degeneracy
persists. Introduce the isotropy subgroup $G_{x^{*}}$, consisting
of all those $g\in G$ which satisfy $gx^{*}=x^{*}$. Since $f$ is
$G$-equivariant, we have
\[
f\left(x^{*}\right)=\gamma\left(g\right)\circ f\left(x^{*}\right)\circ\gamma\left(g^{-1}\right)
\]
for each $g\in G_{x^{*}}$. Therefore $G_{x^{*}}$ sends $f\left(x^{*}\right)$
to itself. In turn, the corresponding tangent map sends the tangent
space $T_{f\left(x^{*}\right)}M$ to itself. In this way $T_{f\left(x^{*}\right)}M$
becomes an orthogonal representation of $G_{x^{*}}$, which we write
as $W=T_{f\left(x^{*}\right)}M$. Note that $W$ contains $T_{f\left(x^{*}\right)}D$
as a $G_{x^{*}}$-invariant subspace, since $D$ is $G$-invariant.
Therefore we can decompose $W$ as $W=W_{D}\oplus W_{D}^{\perp}$,
where $W_{D}=T_{f\left(x^{*}\right)}D$ and where $W_{D}^{\perp}$
is the orthogonal complement of $T_{f\left(x^{*}\right)}D$ in $T_{f\left(x^{*}\right)}M$. 

We will be most interested in the space $W_{D}^{\perp}$. This is
a two-dimensional representation of $G_{x^{*}}$ which describes tangent
directions in $M$ at $f\left(x^{*}\right)\in D$ which are transverse
to the submanifold $D$, that is, directions which break the double
degeneracy. It is straightforward to work out an explicit formula
for the character associated with the representation $W_{D}^{\perp}$,
which we now describe. Since $f\left(x^{*}\right)\in D$, the lowest
eigenspace $\ker\left(f\left(x^{*}\right)-\lambda_{1}\left(x^{*}\right)\right)$
is a two-dimensional subspace of $\mathcal{H}$. Moreover, $f\left(x^{*}\right)$
commutes with $\gamma\left(g\right)$ for each $g\in G_{x^{*}}$,
so this lowest eigenspace is a representation of $G_{x^{*}}$ with
some associated character $\chi$. Then the character associated with
the representation $W_{D}^{\perp}$ is given by 
\[
\chi_{W_{D}^{\perp}}\left(g\right)=\frac{1}{2}\left(\chi\left(g\right)^{2}+\chi\left(g^{2}\right)\right)-1
\]
for each $g\in G_{x^{*}}$.

In the previous section (no symmetry), we chose to identify an open
neighbourhood of $f\left(x^{*}\right)\in M$ with an open neighbourhood
of $\left(0,0\right)\in\mathbb{R}^{N-2}\times\mathbb{R}^{2}$, with
$D$ corresponding to $\mathbb{R}^{N-2}\times\left\{ 0\right\} $.
This could be described as working in local coordinates adapted to
the submanifold $D$. We can do the same here, but we also have the
action of $G_{x^{*}}$ to think about. It turns out that we can identify
a $G_{x^{*}}$-invariant open neighbourhood of $f\left(x^{*}\right)\in M$
with a $G_{x^{*}}$-invariant open neighbourhood of $\left(0,0\right)\in W_{D}\oplus W_{D}^{\perp}$,
with $D$ corresponding to $W_{D}\times\left\{ 0\right\} $. We can
do this by means of a smooth $G_{x^{*}}$-equivariant map. The point
is that these are coordinates in which the submanifold $D$ takes
a simple form, and at the same time the action of $G_{x^{*}}$ is
\emph{linear}, with $G_{x^{*}}$ acting exactly like the representation
$W_{D}\oplus W_{D}^{\perp}$. The proof of the existence of these
coordinates uses Bochner's linearization theorem, and is explained
in detail by Field \cite{Fieldlectures}.

Now we consider $x^{*}\in P$. By definition, $G_{x^{*}}$ sends $x^{*}$
to itself. Therefore $T_{x^{*}}P$ is an orthogonal representation
of $G_{x^{*}}$. Define $G\left(x^{*}\right)$ to be the $G$-orbit
through $x^{*}$. It is clear that $G\left(x^{*}\right)$ is $G_{x^{*}}$-invariant,
so $T_{x^{*}}G\left(x^{*}\right)$ forms a $G_{x^{*}}$-invariant
subspace of $T_{x^{*}}P$. This allows us to decompose $T_{x^{*}}P$
as 
\[
T_{x^{*}}P=T_{x^{*}}G\left(x^{*}\right)\oplus T_{x^{*}}G\left(x^{*}\right)^{\perp},
\]
where $T_{x^{*}}G\left(x^{*}\right)^{\perp}$ is the orthogonal complement
of $T_{x^{*}}G\left(x^{*}\right)$ in $T_{x^{*}}P$. Write $V=T_{x^{*}}G\left(x^{*}\right)^{\perp}$
for this representation of $G_{x^{*}}$.

Choose a slice $S_{x^{*}}$ for the $G$-action at $x^{*}$, that
is, a submanifold which is transverse to the orbit $G\left(x^{*}\right)$
at $x^{*}$ and invariant under $G_{x^{*}}$. From now on we restrict
$f$ to $S_{x^{*}}$, since this restriction determines the behaviour
of $f$ on the whole of $G\left(S_{x^{*}}\right)$, a neighbourhood
of $G(x^{*})$, by equivariance of $f$. Focusing now on $S_{x^{*}}$,
we identify a $G_{x^{*}}$-invariant open neighbourhood of $x^{*}\in S_{x^{*}}$
with a $G_{x^{*}}$-invariant open neighbourhood $\mathcal{U}$ of
the origin $0\in V$ by means of a smooth $G_{x^{*}}$-equivariant
map. Then the local behaviour of $f$ on $S_{x^{*}}$ is expressed
by a map
\[
\tilde{f}:\mathcal{U}\subset V\to W_{D}\oplus W_{D}^{\perp}.
\]
Define $\pi:W_{D}\oplus W_{D}^{\perp}\to W_{D}^{\perp}$ to be projection
onto the second factor, and $F=\pi\circ\tilde{f}$, so that
\[
F:\mathcal{U}\subset V\to W_{D}^{\perp}.
\]
Note that $F\left(0\right)=0$ and, locally, $\Sigma\cap S_{x^{*}}$
is given by solutions of the equation $F=0$. $F$ is a $G_{x^{*}}$-equivariant
smooth map from $\mathcal{U}\subset V$ to $W_{D}^{\perp}$, where
$V$ and $W_{D}^{\perp}$ are representations of $G_{x^{*}}$. 

Generic conditions which characterise the solution set of $F=0$ for
$F$ an equivariant smooth map between representations have been given
by Field and Bierstone. Following their work, we assume that the map
$F$ is \emph{$G_{x^{*}}$-transverse} to $0\in W_{D}^{\perp}$. Given
this transversality assumption, it is possible to characterise the
solution set of $F=0$. The results depend on the group $G_{x^{*}}$
involved, and the isomorphism types of the representations $V$ and
$W_{D}^{\perp}$. We will not give an exhaustive description (for
all groups and representations), but illustrate the approach with
some particular examples relevant to molecules. We end this section
with a brief summary of the definition of equivariant transversality,
followed by a simple example to illustrate the definition and our
approach. For full details on the theory of equivariant transversality,
see Field \cite{Fieldlectures} and Bierstone \cite{Bierstone}. 

\subsubsection*{Equivariant transversality}

Suppose $\mathcal{G}$ is a compact Lie group, and that
\[
\mathcal{F}:\mathcal{V}\to\mathcal{W}
\]
is a $\mathcal{G}$-equivariant smooth map between $\mathcal{G}$-representations
$\mathcal{V}$ and $\mathcal{W}$ with $\mathcal{F}\left(0\right)=0$.

Let $\mathcal{F}_{1},\ldots,\mathcal{F}_{r}$ be a finite set of polynomial
generators for the module $C_{\mathcal{G}}^{\infty}\left(\mathcal{V},\mathcal{W}\right)$
of $\mathcal{G}$-equivariant smooth maps, over the ring $C_{\mathcal{G}}^{\infty}\left(\mathcal{V}\right)$
of $\mathcal{G}$-invariant smooth functions on $\mathcal{V}$. There
are $\mathcal{G}$-invariant smooth functions $h_{1},\ldots,h_{r}$
such that
\[
\mathcal{F}\left(x\right)=\sum_{i=1}^{r}h_{i}\left(x\right)\mathcal{F}_{i}\left(x\right).
\]
This means that we can factor $\mathcal{F}$ as
\[
\mathcal{F}=U\circ\Gamma_{h},
\]
where 
\[
\Gamma_{h}:\mathcal{V}\to\mathcal{V}\times\mathbb{R}^{r}
\]
\[
x\mapsto\left(x,h\left(x\right)\right)
\]
is the graph of $h\left(x\right)=\left(h_{1}\left(x\right),\ldots,h_{r}\left(x\right)\right)$,
and where
\[
U:\mathcal{V}\times\mathbb{R}^{r}\to\mathcal{W}
\]
\[
U\left(x,h\right)=\sum_{i=1}^{r}h_{i}\mathcal{F}_{i}\left(x\right).
\]
Note that, since $\mathcal{F}\left(0\right)=0$, we have $\Gamma_{h}\left(0\right)\in U^{-1}\left(0\right)$.
We say that $\mathcal{F}$ is $\mathcal{G}$-transverse to $0\in\mathcal{W}$
at $0\in\mathcal{V}$ if the map $\Gamma_{h}$ is transverse to $U^{-1}\left(0\right)$
at $0\in\mathcal{V}$.

\subsubsection*{Example}

Suppose $\mathcal{G}=\mathbb{Z}_{2}$, $\mathcal{V}=2A_{1}\oplus A_{2}$
and $\mathcal{W}=A_{1}\oplus A_{2}$. Here $A_{1}$ and $A_{2}$ are
the trivial and non-trivial irreducible representations of $\mathbb{Z}_{2}$. 

Pick bases for $\mathcal{V}$ and $\mathcal{W}$ so that the action
of the non-trivial element of $\mathbb{Z}_{2}$ is given by
\[
\begin{pmatrix}x,y,z\end{pmatrix}\mapsto\begin{pmatrix}x,y,-z\end{pmatrix}
\]
on $\mathcal{V}$, and 
\[
\begin{pmatrix}a,b\end{pmatrix}\mapsto\begin{pmatrix}a,-b\end{pmatrix}
\]
on $\mathcal{W}$. 

A set of polynomial generators for $C_{\mathbb{Z}_{2}}^{\infty}\left(\mathcal{V},\mathcal{W}\right)$
over the ring $C_{\mathbb{Z}_{2}}^{\infty}\left(\mathcal{V}\right)$
is given by 
\[
\mathcal{F}_{1}\left(x,y,z\right)=\begin{pmatrix}1,0\end{pmatrix},\hspace{1em}\mathcal{F}_{2}\left(x,y,z\right)=\begin{pmatrix}0,z\end{pmatrix}.
\]
Any $\mathbb{Z}_{2}$-equivariant smooth map 
\[
\mathcal{F}:\mathcal{V}\to\mathcal{W}
\]
with $\mathcal{F}\left(0\right)=0$ can be written as 
\[
\mathcal{F}\left(x,y,z\right)=h_{1}\left(x,y,z\right)\begin{pmatrix}1,0\end{pmatrix}+h_{2}\left(x,y,z\right)\begin{pmatrix}0,z\end{pmatrix},
\]
where $h_{1}$ and $h_{2}$ are $\mathbb{Z}_{2}$-invariant smooth
functions. We factor this as
\[
\mathcal{F}=U\circ\Gamma_{h},
\]
where
\[
\Gamma_{h}:\mathcal{V}\to\mathcal{V}\times\mathbb{R}^{2}
\]
\[
\left(x,y,z\right)\mapsto\left(x,y,z,h_{1}\left(x,y,z\right),h_{2}\left(x,y,z\right)\right)
\]
is the graph of $h\left(x,y,z\right)=\left(h_{1}\left(x,y,z\right),h_{2}\left(x,y,z\right)\right)$,
and where
\[
U:\mathcal{V}\times\mathbb{R}^{2}\to\mathcal{W}
\]
\[
U\left(x,y,z,h_{1},h_{2}\right)=h_{1}\begin{pmatrix}1,0\end{pmatrix}+h_{2}\begin{pmatrix}0,z\end{pmatrix}.
\]
Note that, since $\mathcal{F}\left(0\right)=0$, we have $\Gamma_{h}\left(0\right)\in U^{-1}\left(0\right)$,
so
\[
\Gamma_{h}\left(0,0,0\right)=\left(0,0,0,0,h_{2}\left(0,0,0\right)\right).
\]
 The set $U^{-1}\left(0\right)$ is given explicitly in this case
by
\[
h_{1}=h_{2}z=0.
\]

We say that $\mathcal{F}$ is $\mathcal{\mathbb{Z}}_{2}$-transverse
to $0\in\mathcal{W}$ at $0\in\mathcal{V}$ if the map $\Gamma_{h}$
is transverse to $U^{-1}\left(0\right)$ at $0\in\mathcal{V}$.

We consider the two possible cases, $h_{2}\left(0,0,0\right)\neq0$
or $h_{2}\left(0,0,0\right)=0$:
\begin{itemize}
\item if $h_{2}\left(0,0,0\right)\neq0$, then transversality to $U^{-1}\left(0\right)$
implies $h_{1x}\left(0,0,0\right)\neq0$ (possibly after swapping
$x,y$). This condition on the partial derivative $h_{1x}$ means
we can make the $\mathbb{Z}_{2}$-equivariant coordinate change
\[
\tilde{x}=h_{1}\left(x,y,z\right),\hspace{1em}\tilde{y}=y,\hspace{1em}\tilde{z}=h_{2}\left(x,y,z\right)z
\]
in a neighbourhood of the origin. This coordinate change (and a subsequent
relabeling $\tilde{x}\to x$, $\tilde{y}\to y$, $\tilde{z}\to z$)
brings $\mathcal{F}$ to the form 
\[
\mathcal{F}:\begin{pmatrix}x,y,z\end{pmatrix}\mapsto\begin{pmatrix}x,z\end{pmatrix}.
\]
We see that 
\[
\mathcal{F}^{-1}\left(0\right)=\left\{ x=z=0\right\} ,
\]
so the solutions of $\mathcal{F}=0$ form a $\mathbb{Z}_{2}$-invariant
smooth curve. Note that all the points $x\in\mathcal{F}^{-1}\left(0\right)$
have $z=0$, so have isotropy group $\mathbb{Z}_{2}$.
\item if $h_{2}\left(0,0,0\right)=0$, then transversality to $U^{-1}\left(0\right)$
implies 
\[
\begin{pmatrix}h_{1x} & h_{1y}\\
h_{2x} & h_{2y}
\end{pmatrix}
\]
is of full rank at $\left(0,0,0\right)$. This condition on the partial
derivatives of $h_{1}$ and $h_{2}$ means we can make the $\mathbb{Z}_{2}$-equivariant
coordinate change
\[
\tilde{x}=h_{1}\left(x,y,z\right),\hspace{1em}\tilde{y}=h_{2}\left(x,y,z\right),\hspace{1em}\tilde{z}=z
\]
in a neighbourhood of the origin. This coordinate change brings $\mathcal{F}$
to the form 
\[
\mathcal{F}:\begin{pmatrix}x,y,z\end{pmatrix}\mapsto\begin{pmatrix}x,yz\end{pmatrix}.
\]
We see that 
\[
\mathcal{F}^{-1}\left(0\right)=\left\{ x=z=0\right\} \cup\left\{ x=y=0\right\} 
\]
so the solutions of $\mathcal{F}=0$ are given by a union of two $\mathbb{Z}_{2}$-invariant
smooth curves. The first curve, defined by $x=z=0$, consists of points
with isotropy group $\mathbb{Z}_{2}$. The second curve branches off
this first curve at the origin, and consists of points with trivial
isotropy group. The set $\mathcal{F}^{-1}\left(0\right)$ is a submanifold
of codimension $2$ everywhere apart from at the origin.
\end{itemize}

\section{Application to Born-Oppenheimer theory}

We are interested in Born-Oppenheimer theory for polyatomic molecules,
in which the electronic Hamiltonian can be viewed as a smooth map
from nuclear configuration space to the space of self-adjoint operators
on the electronic Hilbert space $\mathcal{H}_{e}$. We ignore electronic
spin and relativistic effects (such as spin-orbit coupling). In such
circumstances we can view $\mathcal{H}_{e}$ (which is nominally a
complex vector space) as the complexification of a real vector space,
with all operators of interest (symmetries, the electronic Hamiltonian)
acting on this real vector space and time-reversal corresponding to
complex conjugation. Time-reversal symmetry is dealt with by restricting
our attention to this underlying real vector space from the outset.
Henceforth we simply use $\mathcal{H}_{e}$ to denote this real vector
space. In a real molecular system, $\mathcal{H}_{e}$ is infinite-dimensional,
which is in contrast to the vector space $\mathcal{H}$ considered
so far. Nevertheless, in this section we naively apply the above formalism
to molecules.

We identify $P$ with the nuclear configuration space. That is, a
point $x\in P$ specifies the spatial position of all $\mathcal{N}$
atomic nuclei:
\[
\left(\mathbf{X}_{1},\mathbf{X}_{2},\ldots,\mathbf{X}_{\mathcal{N}}\right),\hspace{1em}\mathbf{X}_{\alpha}\in\mathbb{R}^{3}\hspace{1em}\left(\alpha=1,\ldots,\mathcal{N}\right).
\]
We eliminate the overall translational degrees of freedom of the molecule
by assuming that 
\[
\sum_{\alpha=1}^{\mathcal{N}}\mathbf{X}_{\alpha}=0,
\]
which is three conditions, so that $P$ is $\left(3\mathcal{N}-3\right)$-dimensional.
We also equip $P$ with an inner product according to 
\[
\langle x,y\rangle=\sum_{\alpha=1}^{\mathcal{N}}\mathbf{X}_{\alpha}\cdot\mathbf{Y}_{\alpha}.
\]

The relevant group is $G=O\left(3\right)\times\mathcal{S}$, the product
of $O\left(3\right)$ (which acts on the spatial positions of all
the particles) and the group $\mathcal{S}$ of permutations of atoms
with identical nuclear charges. 

A general element $\left(R,\pi\right)\in G$ acts on $P$ by isometries
according to
\[
\mathbf{X}_{\alpha}\to R\mathbf{X}_{\pi^{-1}\left(\alpha\right)}\hspace{1em}\left(\alpha=1,\ldots,\mathcal{N}\right).
\]
On the other hand, $\left(R,\pi\right)\in G$ acts on $\mathcal{H}_{e}$
by orthogonal linear maps, 
\[
\gamma\left(R\right):\mathcal{H}_{e}\to\mathcal{H}_{e}
\]
which implement the spatial transformations $R\in O\left(3\right)$
on the electronic Hilbert space $\mathcal{H}_{e}$, with the $\pi\in\mathcal{S}$
acting trivially on $\mathcal{H}_{e}$. 

We take the electronic Hamiltonian to consist of the electronic kinetic
energy, plus Coulomb interactions (between all charged particles -
electrons and atomic nuclei). In particular, the electronic Hamiltonian
is invariant under simultaneous $O\left(3\right)$ transformations
of spatial positions of all nuclei and electrons, and invariant under
$\mathcal{S}$. Another way of saying this is that the electronic
Hamiltonian, regarded as a map 
\[
f:P\to M,
\]
is a $G$-equivariant smooth map.

Suppose $x^{*}\in P$ satisfies $f\left(x^{*}\right)\in D$. Then,
as in the previous section, we have the $G_{x^{*}}$-representations
$V=T_{x^{*}}G\left(x^{*}\right)^{\perp}$ and $W_{D}^{\perp}=T_{f\left(x^{*}\right)}D^{\perp}$.
We assume for simplicity that $x^{*}$ is a noncolinear configuration
of the molecule. In that case, it is easily shown that the representation
$V$ has associated character given by 
\[
\chi_{V}\left(R,\pi\right)=\left({\rm tr}R\right)\left({\rm fix}\left(\pi\right)-\det R-1\right),
\]
for each $\left(R,\pi\right)\in G_{x^{*}}$, where ${\rm fix}\left(\pi\right)$
means the number of atoms fixed by the permutation $\pi$. On the
other hand, for the representation $W_{D}^{\perp}$, we repeat our
general formula from earlier:
\[
\chi_{W_{D}^{\perp}}\left(R,\pi\right)=\frac{1}{2}\left(\left[\chi\left(R,\pi\right)\right]^{2}+\chi\left(R^{2},\pi^{2}\right)\right)-1
\]
where $\chi$ is the character associated to the action of $G_{x^{*}}$
on the doubly-degenerate ground state eigenspace of $f\left(x^{*}\right)$.

\subsubsection*{Example $1$: $C_{2v}$ symmetry in triatomics}

Let $x^{*}\in P$ denote an isosceles triangular configuration of
a triatomic molecule, with atom $1$ at the apex of the triangle and
atoms $2$ and $3$ having identical nuclear charges. Then, assuming
the triangle is oriented to lie in the $X=0$ plane with the symmetry
axis in the $Z$-direction, we have that $G_{x^{*}}$ is generated
by 
\[
\left(R_{\pi}\left(\hat{Z}\right),\left(23\right)\right),
\]
that is, rotation by $\pi$ about the $Z$-axis while swapping atoms
$2$ and $3$, and
\[
\left(-R_{\pi}\left(\hat{X}\right),e\right),
\]
that is, reflection in the plane of the molecule.

The projection 
\[
O\left(3\right)\times\mathcal{S}\to O\left(3\right)
\]
onto the first factor gives rise to an isomorphism 
\[
G_{x^{*}}\cong C_{2v}\leq O\left(3\right)
\]
which allows us to label elements of $G_{x^{*}}$ by elements of $C_{2v}$.
We use the notation $C_{2}\left(Z\right)$, $\sigma_{v}\left(YZ\right)$,
standard in the chemical literature, for the relevant generators of
$C_{2v}$.

Using character theory we can identify 
\[
V\cong2A_{1}\oplus B_{2},
\]
corresponding physically to the breathing ($A_{1}$) mode, and the
symmetric ($A_{1}$)/ asymmetric ($A_{2}$) stretching modes of the
isosceles triangle. We pick a basis for $V$, with respect to which
we have the actions
\[
C_{2}\left(Z\right):\left(x,y,z\right)\mapsto\left(x,y,-z\right),\hspace{1em}\sigma_{v}\left(YZ\right):\left(x,y,z\right)\mapsto\left(x,y,z\right).
\]

Suppose $f\left(x^{*}\right)\in D$. The representation $W_{D}^{\perp}$
depends on the representation of $C_{2v}$ on the doubly-degenerate
ground state eigenspace of $f\left(x^{*}\right)$. We have four possibilities:
\[
\left(i\right)\hspace{1em}W_{D}^{\perp}\cong2A_{1}\hspace{1em}{\rm if}\hspace{1em}\ker\left(f\left(x^{*}\right)-E_{0}\left(x^{*}\right)\right)\cong2A_{1},2A_{2},2B_{1},2B_{2},
\]
\[
\left(ii\right)\hspace{1em}W_{D}^{\perp}\cong A_{1}\oplus A_{2}\hspace{1em}{\rm if}\hspace{1em}\ker\left(f\left(x^{*}\right)-E_{0}\left(x^{*}\right)\right)\cong A_{1}\oplus A_{2},B_{1}\oplus B_{2},
\]
\[
\left(iii\right)\hspace{1em}W_{D}^{\perp}\cong A_{1}\oplus B_{1}\hspace{1em}{\rm if}\hspace{1em}\ker\left(f\left(x^{*}\right)-E_{0}\left(x^{*}\right)\right)\cong A_{1}\oplus B_{1},A_{2}\oplus B_{2},
\]
\[
\left(iv\right)\hspace{1em}W_{D}^{\perp}\cong A_{1}\oplus B_{2}\hspace{1em}{\rm if}\hspace{1em}\ker\left(f\left(x^{*}\right)-E_{0}\left(x^{*}\right)\right)\cong A_{1}\oplus B_{2},A_{2}\oplus B_{1}.
\]
We consider each of these in turn. 

\subsubsection*{Case $(i)$ }

Pick a basis for $W_{D}^{\perp}\cong2A_{1}$ so that we have the actions
\[
C_{2}\left(Z\right):\left(a,b\right)\mapsto\left(a,b\right),\hspace{1em}\sigma_{v}\left(YZ\right):\left(a,b\right)\mapsto\left(a,b\right).
\]
A set of polynomial generators for $C_{C_{2v}}^{\infty}\left(V,W_{D}^{\perp}\right)$
over the ring $C_{C_{2v}}^{\infty}\left(V\right)$ is given by 
\[
\mathcal{F}_{1}\left(x,y,z\right)=\begin{pmatrix}1,0\end{pmatrix},\hspace{1em}\mathcal{F}_{2}\left(x,y,z\right)=\begin{pmatrix}0,1\end{pmatrix}.
\]
Any $C_{2v}$-equivariant smooth map 
\[
F:V\to W_{D}^{\perp}
\]
with $F\left(0\right)=0$ can be written as 
\[
F\left(x,y,z\right)=h_{1}\left(x,y,z\right)\begin{pmatrix}1,0\end{pmatrix}+h_{2}\left(x,y,z\right)\begin{pmatrix}0,1\end{pmatrix},
\]
where $h_{1}$ and $h_{2}$ are $C_{2v}$-invariant smooth functions.
We factor this as
\[
F=U\circ\Gamma_{h},
\]
where
\[
\Gamma_{h}:V\to V\times\mathbb{R}^{2}
\]
\[
\left(x,y,z\right)\mapsto\left(x,y,z,h_{1}\left(x,y,z\right),h_{2}\left(x,y,z\right)\right)
\]
is the graph of $h\left(x,y,z\right)=\left(h_{1}\left(x,y,z\right),h_{2}\left(x,y,z\right)\right)$,
and where
\[
U:V\times\mathbb{R}^{2}\to W_{D}^{\perp}
\]
\[
U\left(x,y,z,h_{1},h_{2}\right)=h_{1}\begin{pmatrix}1,0\end{pmatrix}+h_{2}\begin{pmatrix}0,1\end{pmatrix}.
\]
Note that, since $F\left(0\right)=0$, we have $\Gamma_{h}\left(0\right)\in U^{-1}\left(0\right)$,
so
\[
\Gamma_{h}\left(0,0,0\right)=\left(0,0,0,0,0\right).
\]
The set $U^{-1}\left(0\right)$ is given explicitly in this case by
\[
h_{1}=h_{2}=0.
\]
We say that $F$ is $C_{2v}$-transverse to $0\in W_{D}^{\perp}$
at $0\in V$ if the map $\Gamma_{h}$ is transverse to $U^{-1}\left(0\right)$
at $0\in V$. In the present case, this implies 
\[
\begin{pmatrix}h_{1x} & h_{1y} & h_{1z}\\
h_{2x} & h_{2y} & h_{2z}
\end{pmatrix}
\]
is of full rank at $\left(0,0,0\right)$. But $h_{1}$ and $h_{2}$
are $C_{2v}$-invariant functions, so we must have $h_{1z}=h_{2z}=0$
at $\left(0,0,0\right)$. So in fact
\[
\begin{pmatrix}h_{1x} & h_{1y}\\
h_{2x} & h_{2y}
\end{pmatrix}
\]
is of full rank at $\left(0,0,0\right)$. These conditions on the
partial derivatives of $h_{1}$ and $h_{2}$ mean that we can make
the $C_{2v}$-equivariant coordinate change
\[
\tilde{x}=h_{1}\left(x,y,z\right),\hspace{1em}\tilde{y}=h_{2}\left(x,y,z\right),\hspace{1em}\tilde{z}=z
\]
in a neighbourhood of the origin. This coordinate change brings $F$
to the form 
\[
F:\begin{pmatrix}x,y,z\end{pmatrix}\mapsto\begin{pmatrix}x,y\end{pmatrix}.
\]
We see that 
\[
F^{-1}\left(0\right)=\left\{ x=y=0\right\} 
\]
so the solutions of $F=0$ form a $C_{2v}$-invariant smooth curve
through the origin. The origin has isotropy group $C_{2v}$, and the
rest of the curve consists of points with the smaller isotropy group
$C_{s}$.

\subsubsection*{Cases $(ii)$ and $\left(iii\right)$ }

We analyse these cases together as they are very similar. Pick a basis
for $W_{D}^{\perp}$ so that we have the actions
\[
C_{2}\left(Z\right):\left(a,b\right)\mapsto\left(a,\pm b\right),\hspace{1em}\sigma_{v}\left(YZ\right):\left(a,b\right)\mapsto\left(a,-b\right)
\]
where we take $+$ or $-$ in case $(ii)$ and case $(iii)$ respectively.
A set of polynomial generators for $C_{C_{2v}}^{\infty}\left(V,W_{D}^{\perp}\right)$
over the ring $C_{C_{2v}}^{\infty}\left(V\right)$ is given by 
\[
\mathcal{F}_{1}\left(x,y,z\right)=\begin{pmatrix}1,0\end{pmatrix}.
\]
Any $C_{2v}$-equivariant smooth map 
\[
F:V\to W_{D}^{\perp}
\]
with $F\left(0\right)=0$ can be written as 
\[
F\left(x,y,z\right)=h_{1}\left(x,y,z\right)\begin{pmatrix}1,0\end{pmatrix},
\]
where $h_{1}$ is a $C_{2v}$-invariant smooth function. We factor
this as
\[
F=U\circ\Gamma_{h},
\]
where
\[
\Gamma_{h}:V\to V\times\mathbb{R}
\]
\[
\left(x,y,z\right)\mapsto\left(x,y,z,h_{1}\left(x,y,z\right)\right)
\]
is the graph of $h\left(x,y,z\right)=\left(h_{1}\left(x,y,z\right)\right)$,
and where
\[
U:V\times\mathbb{R}\to W_{D}^{\perp}
\]
\[
U\left(x,y,z,h_{1}\right)=h_{1}\begin{pmatrix}1,0\end{pmatrix}.
\]
Note that, since $F\left(0\right)=0$, we have $\Gamma_{h}\left(0\right)\in U^{-1}\left(0\right)$,
so
\[
\Gamma_{h}\left(0,0,0\right)=\left(0,0,0,0\right).
\]
The set $U^{-1}\left(0\right)$ is given explicitly in this case by
\[
h_{1}=0.
\]
We say that $F$ is $C_{2v}$-transverse to $0\in W_{D^{\perp}}$
at $0\in V$ if the map $\Gamma_{h}$ is transverse to $U^{-1}\left(0\right)$
at $0\in V$. In the present case, this implies 
\[
\begin{pmatrix}h_{1x} & h_{1y} & h_{1z}\end{pmatrix}
\]
is nonvanishing at $\left(0,0,0\right)$. But $h_{1}$ is a $C_{2v}$-invariant
function, so we must have $h_{1z}=0$ at $\left(0,0,0\right)$. So
in fact
\[
\begin{pmatrix}h_{1x} & h_{1y}\end{pmatrix}
\]
is nonvanishing at $\left(0,0,0\right)$. This condition on the partial
derivatives of $h_{1}$ means that we can make the $C_{2v}$-equivariant
coordinate change (possibly after swapping $x,y$)
\[
\tilde{x}=h_{1}\left(x,y,z\right),\hspace{1em}\tilde{y}=y,\hspace{1em}\tilde{z}=z
\]
in a neighbourhood of the origin. This coordinate change brings $F$
to the form 
\[
F:\begin{pmatrix}x,y,z\end{pmatrix}\mapsto\begin{pmatrix}x,0\end{pmatrix}.
\]
We see that 
\[
F^{-1}\left(0\right)=\left\{ x=0\right\} 
\]
so the solutions of $F=0$ form a $C_{2v}$-invariant smooth surface.
On this surface, the $C_{2v}$-invariant smooth curve $x=z=0$ consists
of points with isotropy group $C_{2v}$. The rest of the surface consists
of points with isotropy group $C_{s}$.

\subsubsection*{Case $(iv)$ }

Pick a basis for $W_{D}^{\perp}$ so that we have the actions
\[
C_{2}\left(Z\right):\left(a,b\right)\mapsto\left(a,-b\right),\hspace{1em}\sigma_{v}\left(YZ\right):\left(a,b\right)\mapsto\left(a,b\right).
\]
A set of polynomial generators for $C_{C_{2v}}^{\infty}\left(V,W_{D}^{\perp}\right)$
over the ring $C_{C_{2v}}^{\infty}\left(V\right)$ is given by 
\[
\mathcal{F}_{1}\left(x,y,z\right)=\begin{pmatrix}1,0\end{pmatrix},\hspace{1em}\mathcal{F}_{2}\left(x,y,z\right)=\begin{pmatrix}0,z\end{pmatrix}.
\]
Any $C_{2v}$-equivariant smooth map 
\[
F:V\to W_{D}^{\perp}
\]
with $F\left(0\right)=0$ can be written as 
\[
F\left(x,y,z\right)=h_{1}\left(x,y,z\right)\begin{pmatrix}1,0\end{pmatrix}+h_{2}\left(x,y,z\right)\begin{pmatrix}0,z\end{pmatrix},
\]
where $h_{1}$ and $h_{2}$ are $C_{2v}$-invariant smooth functions.
We factor this as
\[
F=U\circ\Gamma_{h},
\]
where
\[
\Gamma_{h}:V\to V\times\mathbb{R}^{2}
\]
\[
\left(x,y,z\right)\mapsto\left(x,y,z,h_{1}\left(x,y,z\right),h_{2}\left(x,y,z\right)\right)
\]
is the graph of $h\left(x,y,z\right)=\left(h_{1}\left(x,y,z\right),h_{2}\left(x,y,z\right)\right)$,
and where
\[
U:V\times\mathbb{R}^{2}\to W_{D}^{\perp}
\]
\[
U\left(x,y,z,h_{1},h_{2}\right)=h_{1}\begin{pmatrix}1,0\end{pmatrix}+h_{2}\begin{pmatrix}0,z\end{pmatrix}.
\]
Note that, since $F\left(0\right)=0$, we have $\Gamma_{h}\left(0\right)\in U^{-1}\left(0\right)$,
so
\[
\Gamma_{h}\left(0,0,0\right)=\left(0,0,0,0,h_{2}\left(0,0,0\right)\right).
\]
The set $U^{-1}\left(0\right)$ is given explicitly in this case by
\[
h_{1}=h_{2}z=0.
\]
We say that $F$ is $C_{2v}$-transverse to $0\in W_{D}^{\perp}$
at $0\in V$ if the map $\Gamma_{h}$ is transverse to $U^{-1}\left(0\right)$
at $0\in V$. We consider the two possible cases, $h_{2}\left(0,0,0\right)\neq0$
or $h_{2}\left(0,0,0\right)=0$:
\begin{itemize}
\item if $h_{2}\left(0,0,0\right)\neq0$, then transversality to $U^{-1}\left(0\right)$
implies $h_{1x}\left(0,0,0\right)\neq0$ (possibly after swapping
$x,y$). This condition on the partial derivative $h_{1x}$ means
we can make the $C_{2v}$-equivariant coordinate change
\[
\tilde{x}=h_{1}\left(x,y,z\right),\hspace{1em}\tilde{y}=y,\hspace{1em}\tilde{z}=h_{2}\left(x,y,z\right)z
\]
in a neighbourhood of the origin. This coordinate change brings $F$
to the form 
\[
F:\begin{pmatrix}x,y,z\end{pmatrix}\mapsto\begin{pmatrix}x,z\end{pmatrix}.
\]
We see that 
\[
F^{-1}\left(0\right)=\left\{ x=z=0\right\} ,
\]
so the solutions of $F=0$ form a $C_{2v}$-invariant smooth curve,
consisting entirely of points with isotropy group $C_{2v}$.
\item if $h_{2}\left(0,0,0\right)=0$, then transversality to $U^{-1}\left(0\right)$
implies 
\[
\begin{pmatrix}h_{1x} & h_{1y}\\
h_{2x} & h_{2y}
\end{pmatrix}
\]
is of full rank at $\left(0,0,0\right)$. This condition on the partial
derivatives of $h_{1}$ and $h_{2}$ means that we can make the $C_{2v}$-equivariant
coordinate change
\[
\tilde{x}=h_{1}\left(x,y,z\right),\hspace{1em}\tilde{y}=h_{2}\left(x,y,z\right),\hspace{1em}\tilde{z}=z
\]
in a neighbourhood of the origin. This coordinate change brings $F$
to the form 
\[
F:\begin{pmatrix}x,y,z\end{pmatrix}\mapsto\begin{pmatrix}x,yz\end{pmatrix}.
\]
We see that 
\[
F^{-1}\left(0\right)=\left\{ x=z=0\right\} \cup\left\{ x=y=0\right\} 
\]
so the solutions of $F=0$ \emph{do not} form a submanifold, but a
\emph{union of two} $C_{2v}$-invariant smooth curves. The first curve,
defined by $x=z=0$, consists entirely of points with isotropy group
$C_{2v}$. The second curve, defined by $x=y=0$, branches off this
first curve and consists of points with isotropy group $C_{s}$. The
set $F^{-1}\left(0\right)$ is a submanifold of codimension $2$ everywhere
apart from at the origin. This is our first example of a \emph{confluence}.
This particular type of confluence has been discovered in numerical
studies of ozone \cite{key-1}, $CH_{2}$ \cite{key-2},
$AlH_{2}$ \cite{key-3}, $BH_{2}$ \cite{key-10}, and further analysed
by Yarkony \cite{key-4}, who describes it as a trifurcation of a
$C_{2v}A-B$ seam of conical intersection. 
\end{itemize}

\subsubsection*{Example $2$: $D_{3h}$ symmetry in triatomics}

Let $x^{*}\in P$ denote an equilateral triangular configuration of
a triatomic molecule, with all atoms having identical nuclear charges.
Then, assuming the triangle is oriented to lie in the $Z=0$ plane,
with the $X$-axis coincident with a $C'_{2}$ symmetry axis which
passes through atom $1$, we have that $G_{x^{*}}$ is generated by
\[
\left(-R_{\pi}\left(\hat{Z}\right),e\right),
\]
 
\[
\left(R_{2\pi/3}\left(\hat{Z}\right),\left(123\right)\right),
\]
and 
\[
\left(R_{\pi}\left(\hat{X}\right),\left(23\right)\right).
\]

The projection 
\[
O\left(3\right)\times\mathcal{S}\to O\left(3\right)
\]
onto the first factor gives rise to an isomorphism 
\[
G_{x^{*}}\cong D_{3h}\leq O\left(3\right)
\]
which allows us to label elements of $G_{x^{*}}$ by elements of $D_{3h}$.

Using character theory we can identify

\[
V\cong A_{1}'\oplus E',
\]
corresponding to the symmetric breathing mode plus a two-dimensional
stretching mode of the equilateral triangle. We pick a basis for $V$,
with respect to which we have the actions

\[
C_{3}\left(Z\right):\begin{pmatrix}x\\
y\\
z
\end{pmatrix}\mapsto\begin{pmatrix}\cos\left(2\pi/3\right) & -\sin\left(2\pi/3\right) & 0\\
\sin\left(2\pi/3\right) & \cos\left(2\pi/3\right) & 0\\
0 & 0 & 1
\end{pmatrix}\begin{pmatrix}x\\
y\\
z
\end{pmatrix},
\]
\[
C_{2}':\begin{pmatrix}x\\
y\\
z
\end{pmatrix}\mapsto\begin{pmatrix}1 & 0 & 0\\
0 & -1 & 0\\
0 & 0 & 1
\end{pmatrix}\begin{pmatrix}x\\
y\\
z
\end{pmatrix},
\]
\[
\sigma_{h}\left(XY\right):\begin{pmatrix}x\\
y\\
z
\end{pmatrix}\mapsto\begin{pmatrix}1 & 0 & 0\\
0 & 1 & 0\\
0 & 0 & 1
\end{pmatrix}\begin{pmatrix}x\\
y\\
z
\end{pmatrix}.
\]

Suppose $f\left(x^{*}\right)\in D$. The representation $W_{D}^{\perp}$
depends on the representation of $D_{3h}$ on the doubly-degenerate
ground state eigenspace of $f\left(x^{*}\right)$. We have five possibilities:
\[
(i)\hspace{1em}W_{D}^{\perp}\cong2A_{1}'\hspace{1em}{\rm if}\hspace{1em}\ker\left(f\left(x^{*}\right)-E_{0}\left(x^{*}\right)\right)\cong2A_{1}',2A_{2}',2A_{1}'',2A_{2}'',
\]
\[
(ii)\hspace{1em}W_{D}^{\perp}\cong A_{1}'\oplus A_{1}''\hspace{1em}{\rm if}\hspace{1em}\ker\left(f\left(x^{*}\right)-E_{0}\left(x^{*}\right)\right)\cong A_{1}'\oplus A_{1}'',A_{2}'\oplus A_{2}'',
\]
\[
(iii)\hspace{1em}W_{D}^{\perp}\cong A_{1}'\oplus A_{2}''\hspace{1em}{\rm if}\hspace{1em}\ker\left(f\left(x^{*}\right)-E_{0}\left(x^{*}\right)\right)\cong A_{1}'\oplus A_{2}'',A_{2}'\oplus A_{1}'',
\]
\[
(iv)\hspace{1em}W_{D}^{\perp}\cong E'\hspace{1em}{\rm if}\hspace{1em}\ker\left(f\left(x^{*}\right)-E_{0}\left(x^{*}\right)\right)\cong E',E'',
\]
\[
(v)\hspace{1em}W_{D}^{\perp}\cong A_{1}'\oplus A_{2}'\hspace{1em}{\rm if}\hspace{1em}\ker\left(f\left(x^{*}\right)-E_{0}\left(x^{*}\right)\right)\cong A_{1}'\oplus A_{2}',A_{1}''\oplus A_{2}''.
\]

We consider each of these in turn. 

\subsubsection*{Case $(i)$ }

Pick a basis for $W_{D}^{\perp}$ so that we have the actions
\[
C_{3}\left(Z\right):\left(a,b\right)\mapsto\left(a,b\right),\hspace{1em}C_{2}':\left(a,b\right)\mapsto\left(a,b\right),\hspace{1em}\sigma_{h}\left(XY\right):\left(a,b\right)\mapsto\left(a,b\right).
\]
A set of polynomial generators for $C_{D_{3h}}^{\infty}\left(V,W_{D}^{\perp}\right)$
over the ring $C_{D_{3h}}^{\infty}\left(V\right)$ is given by 
\[
\mathcal{F}_{1}\left(x,y,z\right)=\begin{pmatrix}1,0\end{pmatrix},\hspace{1em}\mathcal{F}_{2}\left(x,y,z\right)=\begin{pmatrix}0,1\end{pmatrix}.
\]
Any $D_{3h}$-equivariant smooth map 
\[
F:V\to W_{D}^{\perp}
\]
with $F\left(0\right)=0$ can be written as 
\[
F\left(x,y,z\right)=h_{1}\left(x,y,z\right)\begin{pmatrix}1,0\end{pmatrix}+h_{2}\left(x,y,z\right)\begin{pmatrix}0,1\end{pmatrix},
\]
where $h_{1}$ and $h_{2}$ are $D_{3h}$-invariant smooth functions.
We factor this as
\[
F=U\circ\Gamma_{h},
\]
where
\[
\Gamma_{h}:V\to V\times\mathbb{R}^{2}
\]
\[
\left(x,y,z\right)\mapsto\left(x,y,z,h_{1}\left(x,y,z\right),h_{2}\left(x,y,z\right)\right)
\]
is the graph of $h\left(x,y,z\right)=\left(h_{1}\left(x,y,z\right),h_{2}\left(x,y,z\right)\right)$,
and where
\[
U:V\times\mathbb{R}^{2}\to W_{D}^{\perp}
\]
\[
U\left(x,y,z,h_{1},h_{2}\right)=h_{1}\begin{pmatrix}1,0\end{pmatrix}+h_{2}\begin{pmatrix}0,1\end{pmatrix}.
\]
Note that, since $F\left(0\right)=0$, we have $\Gamma_{h}\left(0\right)\in U^{-1}\left(0\right)$,
so
\[
\Gamma_{h}\left(0,0,0\right)=\left(0,0,0,0,0\right).
\]
The set $U^{-1}\left(0\right)$ is given explicitly in this case by
\[
h_{1}=h_{2}=0.
\]
We say that $F$ is $D_{3h}$-transverse to $0\in W_{D}^{\perp}$
at $0\in V$ if the map $\Gamma_{h}$ is transverse to $U^{-1}\left(0\right)$
at $0\in V$. This means that 
\[
\begin{pmatrix}h_{1x} & h_{1y} & h_{1z}\\
h_{2x} & h_{2y} & h_{2z}
\end{pmatrix}
\]
has full rank at $\left(0,0,0\right)$. But $h_{1}$ and $h_{2}$
are $D_{3h}$-invariant functions and, in particular, the first two
columns of this matrix must vanish at $\left(0,0,0\right)$. Such
a matrix has no hope of being full rank. We have reached a contradiction;
degeneracies of this kind generically \emph{do not occur}.

\subsubsection*{Cases $(ii)$ and $\left(iii\right)$}

We analyse these cases together as they are very similar. Pick a basis
for $W_{D}^{\perp}$ so that we have the actions

\[
C_{3}\left(Z\right):\left(a,b\right)\mapsto\left(a,b\right),\hspace{1em}C_{2}':\left(a,b\right)\mapsto\left(a,\pm b\right),\hspace{1em}\sigma_{h}\left(XY\right):\left(a,b\right)\mapsto\left(a,-b\right)
\]
where we take $+$ or $-$ in case $(ii)$ and case $(iii)$ respectively.
A set of polynomial generators for $C_{D_{3h}}^{\infty}\left(V,W_{D}^{\perp}\right)$
over the ring $C_{D_{3h}}^{\infty}\left(V\right)$ is given by 
\[
\mathcal{F}_{1}\left(x,y,z\right)=\begin{pmatrix}1,0\end{pmatrix}.
\]
Any $D_{3h}$-equivariant smooth map 
\[
F:V\to W_{D}^{\perp}
\]
with $F\left(0\right)=0$ can be written as 
\[
F\left(x,y,z\right)=h_{1}\left(x,y,z\right)\begin{pmatrix}1,0\end{pmatrix},
\]
where $h_{1}$ and $h_{2}$ are $D_{3h}$-invariant smooth functions.
We factor this as
\[
F=U\circ\Gamma_{h},
\]
where
\[
\Gamma_{h}:V\to V\times\mathbb{R}^{2}
\]
\[
\left(x,y,z\right)\mapsto\left(x,y,z,h_{1}\left(x,y,z\right)\right)
\]
is the graph of $h\left(x,y,z\right)=\left(h_{1}\left(x,y,z\right)\right)$,
and where
\[
U:V\times\mathbb{R}^{2}\to W_{D}^{\perp}
\]
\[
U\left(x,y,z,h_{1},h_{2}\right)=h_{1}\begin{pmatrix}1,0\end{pmatrix}.
\]
Note that, since $F\left(0\right)=0$, we have $\Gamma_{h}\left(0\right)\in U^{-1}\left(0\right)$,
so
\[
\Gamma_{h}\left(0,0,0\right)=\left(0,0,0,0\right).
\]
The set $U^{-1}\left(0\right)$ is given explicitly in this case by
\[
h_{1}=0.
\]
We say that $F$ is $D_{3h}$-transverse to $0\in W_{D}^{\perp}$
at $0\in V$ if the map $\Gamma_{h}$ is transverse to $U^{-1}\left(0\right)$
at $0\in V$. In this case, transversality to $U^{-1}\left(0\right)$
implies $h_{1z}\neq0$ so we can make the $D_{3h}$-equivariant coordinate
change
\[
\tilde{x}=x,\hspace{1em}\tilde{y}=y,\hspace{1em}\tilde{z}=h_{1}
\]
in a neighbourhood of the origin. This coordinate change brings $F$
to the form 
\[
F:\begin{pmatrix}x,y,z\end{pmatrix}\mapsto\begin{pmatrix}z,0\end{pmatrix}.
\]
We see that 
\[
F^{-1}\left(0\right)=\left\{ z=0\right\} ,
\]
so the solutions of $F=0$ form a $D_{3h}$-invariant smooth surface
through the origin. The origin has isotropy group $D_{3h}$. The surface
contains three smooth curves through the origin consisting of points
with isotropy group $C_{2v}$ (away from the origin). The rest of
the surface consists of points with isotropy group $C_{s}$.

\subsubsection*{Case $(iv)$ }

Pick a basis for $W_{D}^{\perp}$ so that we have the actions

\[
C_{3}\left(Z\right):\begin{pmatrix}a\\
b
\end{pmatrix}\mapsto\begin{pmatrix}\cos\left(2\pi/3\right) & -\sin\left(2\pi/3\right)\\
\sin\left(2\pi/3\right) & \cos\left(2\pi/3\right)
\end{pmatrix}\begin{pmatrix}a\\
b
\end{pmatrix},
\]
\[
C_{2}':\begin{pmatrix}a\\
b
\end{pmatrix}\mapsto\begin{pmatrix}1 & 0\\
0 & -1
\end{pmatrix}\begin{pmatrix}a\\
b
\end{pmatrix},
\]
\[
\sigma_{h}\left(XY\right):\begin{pmatrix}a\\
b
\end{pmatrix}\mapsto\begin{pmatrix}1 & 0\\
0 & 1
\end{pmatrix}\begin{pmatrix}a\\
b
\end{pmatrix}.
\]
A set of polynomial generators for $C_{D_{3h}}^{\infty}\left(V,W_{D}^{\perp}\right)$
over the ring $C_{D_{3h}}^{\infty}\left(V\right)$ is given by 
\[
\mathcal{F}_{1}\left(x,y,z\right)=\begin{pmatrix}x,y\end{pmatrix},\hspace{1em}\mathcal{F}_{2}\left(x,y,z\right)=\begin{pmatrix}x^{2}-y^{2},-2xy\end{pmatrix}.
\]
Any $D_{3h}$-equivariant smooth map 
\[
F:V\to W_{D}^{\perp}
\]
with $F\left(0\right)=0$ can be written as 
\[
F\left(x,y,z\right)=h_{1}\left(x,y,z\right)\begin{pmatrix}x,y\end{pmatrix}+h_{2}\left(x,y,z\right)\begin{pmatrix}x^{2}-y^{2},-2xy\end{pmatrix},
\]
where $h_{1}$ and $h_{2}$ are $D_{3h}$-invariant smooth functions.
We factor this as
\[
F=U\circ\Gamma_{h},
\]
where
\[
\Gamma_{h}:V\to V\times\mathbb{R}^{2}
\]
\[
\left(x,y,z\right)\mapsto\left(x,y,z,h_{1}\left(x,y,z\right),h_{2}\left(x,y,z\right)\right)
\]
is the graph of $h\left(x,y,z\right)=\left(h_{1}\left(x,y,z\right),h_{2}\left(x,y,z\right)\right)$,
and where
\[
U:V\times\mathbb{R}^{2}\to W_{D}^{\perp}
\]
\[
U\left(x,y,z,h_{1},h_{2}\right)=h_{1}\begin{pmatrix}x,y\end{pmatrix}+h_{2}\begin{pmatrix}x^{2}-y^{2},-2xy\end{pmatrix}.
\]
Note that, since $F\left(0\right)=0$, we have $\Gamma_{h}\left(0\right)\in U^{-1}\left(0\right)$,
so
\[
\Gamma_{h}\left(0,0,0\right)=\left(0,0,0,h_{1}\left(0,0,0\right),h_{2}\left(0,0,0\right)\right).
\]
The set $U^{-1}\left(0\right)$ is given explicitly in this case by
\[
h_{1}x+h_{2}\left(x^{2}-y^{2}\right)=h_{1}y-2h_{2}xy=0.
\]
Note that this can equivalently be written as
\[
\left\{ y-\sqrt{3}x=h_{1}-2h_{2}x=0\right\} \cup\left\{ y+\sqrt{3}x=h_{1}-2h_{2}x=0\right\} 
\]
\[
\cup\left\{ h_{2}=h_{1}=0\right\} \cup\left\{ h_{1}+h_{2}x=y=0\right\} \cup\left\{ x=y=0\right\} .
\]
We say that $F$ is $D_{3h}$-transverse to $0\in W_{D}^{\perp}$
at $0\in V$ if the map $\Gamma_{h}$ is transverse to $U^{-1}\left(0\right)$
at $0\in V$. We consider the two cases, $h_{1}\left(0,0,0\right)\neq0$
or $h_{1}\left(0,0,0\right)=0$:
\begin{itemize}
\item if $h_{1}\left(0,0,0\right)\neq0$, we can make the $D_{3h}$-equivariant
coordinate change
\[
\tilde{x}=h_{1}\left(x,y,z\right)x+h_{2}\left(x,y,z\right)\left(x^{2}-y^{2}\right),
\]
\[
\tilde{y}=h_{1}\left(x,y,z\right)y-2h_{2}\left(x,y,z\right)xy,
\]
\[
\tilde{z}=z
\]
in a neighbourhood of the origin. This coordinate change brings $F$
to the form 
\[
F:\begin{pmatrix}x,y,z\end{pmatrix}\mapsto\begin{pmatrix}x,y\end{pmatrix}.
\]
We see that 
\[
F^{-1}\left(0\right)=\left\{ x=y=0\right\} ,
\]
so the solutions of $F=0$ form a $D_{3h}$-invariant smooth curve,
consisting entirely of points with isotropy group $D_{3h}$.
\item if $h_{1}\left(0,0,0\right)=0$, we consider the cases $h_{2}\left(0,0,0\right)=0$
and $h_{2}\left(0,0,0\right)\neq0$. If $h_{2}\left(0,0,0\right)=0$,
then transversality to $U^{-1}\left(0\right)$ is not possible. So
this case does not generically occur. If $h_{2}\left(0,0,0\right)\neq0$,
then transversality to $U^{-1}\left(0\right)$ implies $h_{1z}\left(0,0,0\right)\neq0$.
In this case, we can make the $D_{3h}$-equivariant coordinate change
\[
\tilde{x}=\left|h_{2}\right|^{1/2}x,\hspace{1em}\tilde{y}=\left|h_{2}\right|^{1/2}y,\hspace{1em},\tilde{z}=h_{2}\left|h_{2}\right|^{-3/2}h_{1}
\]
in a neighbourhood of the origin. This coordinate change brings $F$
to the form 
\[
\mathcal{F}:\begin{pmatrix}x,y,z\end{pmatrix}\mapsto\pm\begin{pmatrix}zx+x^{2}-y^{2},zy-2xy\end{pmatrix}
\]
where the $\pm$ corresponds to the sign of $h_{2}\left(0,0,0\right)$.
We see that 
\[
F^{-1}\left(0\right)=\left\{ x=y=0\right\} \cup\left\{ \sqrt{3}x-y=z-2x=0\right\} 
\]
\[
\cup\left\{ \sqrt{3}x+y=z-2x=0\right\} \cup\left\{ z+x=y=0\right\} ,
\]
so the solutions of $F=0$ do not form a submanifold, but a $D_{3h}$-invariant
union of \emph{four} smooth curves. The first curve, defined by $x=y=0$,
consists entirely of points with isotropy group $D_{3h}$. The other
curves branch off this first curve and consist of points with isotropy
group $C_{2v}$. The set $F^{-1}\left(0\right)$ is a submanifold
of codimension $2$ everywhere apart from at the origin. This is our
second example of a confluence. This particular type of confluence
has been discovered in $Li_{3}$ \cite{key-5} and $N_{3}^{+}$ \cite{key-8},
where the authors describe it as a heptafurcation of a symmetry-required
$D_{3h}$ seam of conical intersection.
\end{itemize}

\subsubsection*{Case $(v)$}

Pick a basis for $W_{D}^{\perp}$ so that we have the actions

\[
C_{3}\left(Z\right):\left(a,b\right)\mapsto\left(a,b\right),\hspace{1em}C_{2}':\left(a,b\right)\mapsto\left(a,-b\right),\hspace{1em}\sigma_{h}\left(XY\right):\left(a,b\right)\mapsto\left(a,b\right).
\]
A set of polynomial generators for $C_{D_{3h}}^{\infty}\left(V,W_{D}^{\perp}\right)$
over the ring $C_{D_{3h}}^{\infty}\left(V\right)$ is given by 
\[
\mathcal{F}_{1}\left(x,y,z\right)=\begin{pmatrix}1,0\end{pmatrix},\hspace{1em}\mathcal{F}_{2}\left(x,y,z\right)=\begin{pmatrix}0,y^{3}-3x^{2}y\end{pmatrix}.
\]
Any $D_{3h}$-equivariant smooth map 
\[
F:V\to W_{D}^{\perp}
\]
with $F\left(0\right)=0$ can be written as 
\[
F\left(x,y,z\right)=h_{1}\left(x,y,z\right)\begin{pmatrix}1,0\end{pmatrix}+h_{2}\left(x,y,z\right)\begin{pmatrix}0,y^{3}-3x^{2}y\end{pmatrix},
\]
where $h_{1}$ and $h_{2}$ are $D_{3h}$-invariant smooth functions.
We factor this as
\[
F=U\circ\Gamma_{h},
\]
where
\[
\Gamma_{h}:V\to V\times\mathbb{R}^{2}
\]
\[
\left(x,y,z\right)\mapsto\left(x,y,z,h_{1}\left(x,y,z\right),h_{2}\left(x,y,z\right)\right)
\]
is the graph of $h\left(x,y,z\right)=\left(h_{1}\left(x,y,z\right),h_{2}\left(x,y,z\right)\right)$,
and where
\[
U:V\times\mathbb{R}^{2}\to W_{D}^{\perp}
\]
\[
U\left(x,y,z,h_{1},h_{2}\right)=h_{1}\begin{pmatrix}1,0\end{pmatrix}+h_{2}\begin{pmatrix}0,y^{3}-3x^{2}y\end{pmatrix}.
\]
Note that, since $F\left(0\right)=0$, we have $\Gamma_{h}\left(0\right)\in U^{-1}\left(0\right)$,
so
\[
\Gamma_{h}\left(0,0,0\right)=\left(0,0,0,0,h_{2}\left(0,0,0\right)\right).
\]
The set $U^{-1}\left(0\right)$ is given explicitly in this case by
\[
h_{1}=h_{2}\left(y^{3}-3x^{2}y\right)=0.
\]
Note that this can equivalently be written as
\[
\left\{ h_{1}=h_{2}=0\right\} \cup\left\{ h_{1}=y=0\right\} 
\]
\[
\cup\left\{ h_{1}=y-\sqrt{3}x=0\right\} \cup\left\{ h_{1}=y+\sqrt{3}x=0\right\} .
\]
We say that $F$ is $D_{3h}$-transverse to $0\in W_{D}^{\perp}$
at $0\in V$ if the map $\Gamma_{h}$ is transverse to $U^{-1}\left(0\right)$
at $0\in V$. We consider the two cases, $h_{2}\left(0,0,0\right)\neq0$
or $h_{2}\left(0,0,0\right)=0$. If $h_{2}\left(0,0,0\right)=0$,
then transversality to $U^{-1}\left(0\right)$ is not possible. So
this case does not generically occur. If $h_{2}\left(0,0,0\right)\neq0$,
then transversality to $U^{-1}\left(0\right)$ implies $h_{1z}\left(0,0,0\right)\neq0$
so we can make the $D_{3h}$-equivariant coordinate change
\[
\tilde{x}=xh_{2}^{1/3},\hspace{1em}\tilde{y}=yh_{2}^{1/3},\hspace{1em}\tilde{z}=h_{1}
\]
in a neighbourhood of the origin. This coordinate change brings $F$
to the form 
\[
F:\begin{pmatrix}x,y,z\end{pmatrix}\mapsto\begin{pmatrix}z,y^{3}-3x^{2}y\end{pmatrix}.
\]
We see that 
\[
F^{-1}\left(0\right)=\left\{ z=y=0\right\} \cup\left\{ z=y-\sqrt{3}x=0\right\} \cup\left\{ z=y+\sqrt{3}x=0\right\} 
\]
so the solutions of $F=0$ do not form a submanifold, but a $D_{3h}$-invariant
union of \emph{three} smooth curves. Apart from the origin, which
has isotropy group $D_{3h}$, each of these curves consists of points
with isotropy group $C_{2v}$. The set $F^{-1}\left(0\right)$ is
a submanifold of codimension $2$ everywhere apart from at the origin.
This is our third example of a confluence. This particular type of
confluence has not yet been discovered in a real molecule.

\subsubsection*{Example $3$: $C_{s}$ symmetry at planar configurations of molecules}

Let $x^{*}\in P$ denote a general planar (but noncolinear) configuration
of a molecule with $\mathcal{N}>3$ atoms, oriented so that the molecule
lies in the $Z=0$ plane. We have that $G_{x^{*}}$ is generated by
\[
\left(-R_{\pi}\left(\hat{Z}\right),e\right),
\]
a reflection in the plane of the molecule.

The projection 
\[
O\left(3\right)\times\mathcal{S}\to O\left(3\right)
\]
onto the first factor gives rise to an isomorphism 
\[
G_{x^{*}}\cong C_{s}\leq O\left(3\right)
\]
which allows us to label elements of $G_{x}$ by elements of $C_{s}$.

Using character theory we can identify 
\[
V\cong\left(2\mathcal{N}-3\right)A'\oplus\left(\mathcal{N}-3\right)A''.
\]

We pick a basis for $V$, with respect to which we have the action
\[
\sigma_{h}:\left(x_{1},\ldots,x_{2\mathcal{N}-3},y_{1},\ldots,y_{\mathcal{N}-3}\right)\mapsto\left(x_{1},\ldots,x_{2\mathcal{N}-3},-y_{1},\ldots,-y_{\mathcal{N}-3}\right),
\]
or, written more compactly,
\[
\sigma_{h}:\left(\mathbf{x},\mathbf{y}\right)\mapsto\left(\mathbf{x},-\mathbf{y}\right).
\]

The representation $W_{D}^{\perp}$ depends on the nature of the representation
of $C_{s}$ on the doubly-degenerate ground state eigenspace of $f\left(x^{*}\right)$.
We have two possibilities:
\[
\left(i\right)\hspace{1em}W_{D}^{\perp}\cong2A'\hspace{1em}{\rm if}\hspace{1em}\ker\left(f\left(x^{*}\right)-E_{0}\left(x^{*}\right)\right)\cong2A',2A'',
\]
\[
\left(ii\right)\hspace{1em}W_{D}^{\perp}\cong A'\oplus A''\hspace{1em}{\rm if}\hspace{1em}\ker\left(f\left(x^{*}\right)-E_{0}\left(x^{*}\right)\right)\cong A'\oplus A''.
\]
We consider each of these in turn. 

\subsubsection*{Case $(i)$ }

Pick a basis for $W_{D}^{\perp}$ so that we have the action

\[
C_{s}:\left(a,b\right)\mapsto\left(a,b\right).
\]
A set of polynomial generators for $C_{C_{s}}^{\infty}\left(V,W_{D}^{\perp}\right)$
over the ring $C_{C_{s}}^{\infty}\left(V\right)$ is given by 
\[
\mathcal{F}_{1}\left(\mathbf{x},\mathbf{y}\right)=\begin{pmatrix}1,0\end{pmatrix},\hspace{1em}\mathcal{F}_{2}\left(\mathbf{x},\mathbf{y}\right)=\begin{pmatrix}0,1\end{pmatrix}.
\]
Any $C_{s}$-equivariant smooth map 
\[
F:V\to W_{D}^{\perp}
\]
with $F\left(0\right)=0$ can be written as 
\[
F\left(\mathbf{x},\mathbf{y}\right)=h_{1}\left(\mathbf{x},\mathbf{y}\right)\begin{pmatrix}1,0\end{pmatrix}+h_{2}\left(\mathbf{x},\mathbf{y}\right)\begin{pmatrix}0,1\end{pmatrix},
\]
where $h_{1}$ and $h_{2}$ are $C_{s}$-invariant smooth functions.
We factor this as
\[
F=U\circ\Gamma_{h},
\]
where
\[
\Gamma_{h}:V\to V\times\mathbb{R}^{2}
\]
\[
\left(\mathbf{x},\mathbf{y}\right)\mapsto\left(\mathbf{x},\mathbf{y},h_{1}\left(\mathbf{x},\mathbf{y}\right),h_{2}\left(\mathbf{x},\mathbf{y}\right)\right)
\]
is the graph of $h\left(\mathbf{x},\mathbf{y}\right)=\left(h_{1}\left(\mathbf{x},\mathbf{y}\right),h_{2}\left(\mathbf{x},\mathbf{y}\right)\right)$,
and where
\[
U:V\times\mathbb{R}^{2}\to W_{D}^{\perp}
\]
\[
U\left(\mathbf{x},\mathbf{y},h_{1},h_{2}\right)=h_{1}\begin{pmatrix}1,0\end{pmatrix}+h_{2}\begin{pmatrix}0,1\end{pmatrix}.
\]
Note that, since $F\left(0\right)=0$, we have $\Gamma_{h}\left(0\right)\in U^{-1}\left(0\right)$,
so
\[
\Gamma_{h}\left(\mathbf{0},\mathbf{0}\right)=\left(\mathbf{0},\mathbf{0},0,0\right).
\]
The set $U^{-1}\left(0\right)$ is given explicitly in this case by
\[
h_{1}=h_{2}=0.
\]
We say that $F$ is $C_{s}$-transverse to $0\in W_{D}^{\perp}$ at
$0\in V$ if the map $\Gamma_{h}$ is transverse to $U^{-1}\left(0\right)$
at $0\in V$. This means that 
\[
\begin{pmatrix}h_{1\mathbf{x}} & h_{1\mathbf{y}}\\
h_{2\mathbf{x}} & h_{2\mathbf{y}}
\end{pmatrix}
\]
is of full rank at $\left(\mathbf{x},\mathbf{y}\right)=\left(\mathbf{0},\mathbf{0}\right)$.
Note that $h_{1}$ and $h_{2}$ are $C_{s}$-invariant, so we must
have $h_{1\mathbf{y}}=h_{2\mathbf{y}}=0$. Therefore 
\[
\begin{pmatrix}h_{1\mathbf{x}}\\
h_{2\mathbf{x}}
\end{pmatrix}
\]
is of full rank at $\left(\mathbf{x},\mathbf{y}\right)=\left(\mathbf{0},\mathbf{0}\right)$.
Therefore (after possibly reordering $x_{1},\ldots,x_{2\mathcal{N}-3}$)
we have $h_{1x_{1}}\neq0$ and $h_{2x_{2}}\neq0$. We can make the
$C_{s}$-equivariant coordinate change
\[
\tilde{x}_{1}=h_{1},\hspace{1em}\tilde{x}_{2}=h_{2},\hspace{1em}\tilde{x}_{3}=x_{3},\ldots,\hspace{1em}\tilde{y}_{\mathcal{N}-3}=y_{\mathcal{N}-3}
\]
in a neighbourhood of the origin. This coordinate change brings $F$
to the form
\[
F:\begin{pmatrix}x_{1},x_{2},x_{3},\ldots,y_{\mathcal{N}-3}\end{pmatrix}\mapsto\begin{pmatrix}x_{1},x_{2}\end{pmatrix}.
\]
We see that 
\[
F^{-1}\left(0\right)=\left\{ x_{1}=x_{2}=0\right\} ,
\]
so the solutions of $F=0$ form a $C_{s}$-invariant submanifold of
codimension $2$. Within this submanifold, the points with $\mathbf{y}=\mathbf{0}$
have isotropy group $C_{s}$. The rest consists of points with trivial
isotropy group. 

\subsubsection*{Case $(ii)$ }

Pick a basis for $W_{D}^{\perp}$ so that we have the action

\[
C_{s}:\left(a,b\right)\mapsto\left(a,-b\right).
\]
A set of polynomial generators for $C_{C_{s}}^{\infty}\left(V,W_{D}^{\perp}\right)$
over the ring $C_{C_{s}}^{\infty}\left(V\right)$ is given by 
\[
\mathcal{F}_{1}\left(\mathbf{x},\mathbf{y}\right)=\begin{pmatrix}1,0\end{pmatrix},\hspace{1em}\mathcal{F}_{2}\left(\mathbf{x},\mathbf{y}\right)=\begin{pmatrix}0,y_{1}\end{pmatrix},\ldots,\hspace{1em}\mathcal{F}_{\mathcal{N}-2}\left(\mathbf{x},\mathbf{y}\right)=\begin{pmatrix}0,y_{\mathcal{N}-3}\end{pmatrix}.
\]
Any $C_{s}$-equivariant smooth map 
\[
F:V\to W_{D}^{\perp}
\]
with $F\left(0\right)=0$ can be written as 
\[
F\left(\mathbf{x},\mathbf{y}\right)=h_{1}\left(\mathbf{x},\mathbf{y}\right)\begin{pmatrix}1,0\end{pmatrix}+\sum_{i=2}^{\mathcal{N}-2}h_{i}\left(\mathbf{x},\mathbf{y}\right)\begin{pmatrix}0,y_{i-1}\end{pmatrix},
\]
where $h_{1},\ldots,h_{\mathcal{N}-2}$ are $C_{s}$-invariant smooth
functions. We factor this as
\[
F=U\circ\Gamma_{h},
\]
where
\[
\Gamma_{h}:V\to V\times\mathbb{R}^{\mathcal{N}-2}
\]
\[
\left(\mathbf{x},\mathbf{y}\right)\mapsto\left(\mathbf{x},\mathbf{y},h_{1}\left(\mathbf{x},\mathbf{y}\right),h_{2}\left(\mathbf{x},\mathbf{y}\right),\ldots,h_{\mathcal{N}-2}\left(\mathbf{x},\mathbf{y}\right)\right)
\]
is the graph of $h\left(\mathbf{x},\mathbf{y}\right)=\left(h_{1}\left(\mathbf{x},\mathbf{y}\right),h_{2}\left(\mathbf{x},\mathbf{y}\right),\ldots,h_{\mathcal{N}-2}\left(\mathbf{x},\mathbf{y}\right)\right)$,
and where
\[
U:V\times\mathbb{R}^{\mathcal{N}-2}\to W_{D}^{\perp}
\]
\[
U\left(\mathbf{x},\mathbf{y},h_{1},h_{2},\ldots,h_{N_{n}-2}\right)=h_{1}\begin{pmatrix}1,0\end{pmatrix}+\sum_{i=2}^{\mathcal{N}-2}h_{i}\begin{pmatrix}0,y_{i-1}\end{pmatrix}.
\]
Note that, since $F\left(0\right)=0$, we have $\Gamma_{h}\left(0\right)\in U^{-1}\left(0\right)$,
so
\[
\Gamma_{h}\left(\mathbf{0},\mathbf{0}\right)=\left(\mathbf{0},\mathbf{0},0,h_{2}\left(\mathbf{0},\mathbf{0}\right),\ldots,h_{\mathcal{N}-2}\left(\mathbf{0},\mathbf{0}\right)\right).
\]
The set $U^{-1}\left(0\right)$ is given explicitly in this case by
\[
h_{1}=h_{2}y_{1}=h_{3}y_{2}=\ldots=h_{\mathcal{N}-2}y_{\mathcal{N}-3}=0.
\]
We say that $F$ is $C_{s}$-transverse to $0\in W_{D}^{\perp}$ at
$0\in V$ if the map $\Gamma_{h}$ is transverse to $U^{-1}\left(0\right)$
at $0\in V$. Transversality to $U^{-1}\left(0\right)$ implies $h_{1x_{1}}\left(\mathbf{0},\mathbf{0}\right)\neq0$
(after possibly reordering $x_{1},\ldots,x_{2\mathcal{N}-3}$). Consider
the two possible cases, $h_{2}\left(\mathbf{0},\mathbf{0}\right)\neq0$
(after possibly reordering $y_{1},\ldots,y_{\mathcal{N}-3}$) or $h_{2}\left(\mathbf{0},\mathbf{0}\right)=\ldots=h_{\mathcal{N}-2}\left(\mathbf{0},\mathbf{0}\right)=0$:
\begin{itemize}
\item If $h_{2}\left(\mathbf{0},\mathbf{0}\right)\neq0$, we can make the
$C_{s}$-equivariant coordinate change
\[
\tilde{x}_{1}=h_{1},\hspace{1em}\tilde{x}_{2}=x_{2},\ldots,
\]
\[
\tilde{y}_{1}=\sum_{i=2}^{\mathcal{N}-2}h_{i}y_{i-1},\hspace{1em}\text{\ensuremath{\tilde{y}}}_{2}=y_{2},\ldots,\hspace{1em}\tilde{y}_{\mathcal{N}-3}=y_{\mathcal{N}-3}
\]
in a neighbourhood of the origin. This coordinate change brings $F$
to the form
\[
F:\begin{pmatrix}x_{1},x_{2},x_{3},\ldots,y_{\mathcal{N}-3}\end{pmatrix}\mapsto\begin{pmatrix}x_{1},y_{1}\end{pmatrix}.
\]
We see that 
\[
F^{-1}\left(0\right)=\left\{ x_{1}=y_{1}=0\right\} ,
\]
so the solutions of $F=0$ form a $C_{s}$-invariant submanifold of
codimension $2$. Within this submanifold, the points with $\mathbf{y}=\mathbf{0}$
have isotropy group $C_{s}$. The rest consists of points with trivial
isotropy group. 
\item If $h_{2}\left(\mathbf{0},\mathbf{0}\right)=\ldots=h_{\mathcal{N}-2}\left(\mathbf{0},\mathbf{0}\right)=0$,
then transversality to $U^{-1}\left(0\right)$ implies the matrix
\[
\begin{pmatrix}h_{1x_{1}} & h_{1x_{2}} & \ldots & h_{1x_{2\mathcal{N}-3}}\\
\vdots & \vdots & \vdots & \vdots\\
h_{\mathcal{N}-2,x_{1}} & h_{\mathcal{N}-2,x_{2}} & \ldots & h_{\mathcal{N}-2,x_{2\mathcal{N}-3}}
\end{pmatrix}
\]
is of full rank at $\left(\mathbf{0},\mathbf{0}\right)$. By reordering
$x_{1},\ldots,x_{2N_{n}-3}$ if necessary, we get that the square
submatrix
\[
\begin{pmatrix}h_{1x_{1}} & h_{1x_{2}} & \ldots & h_{1x_{\mathcal{N}-2}}\\
\vdots & \vdots & \vdots & \vdots\\
h_{\mathcal{N}-2,x_{1}} & h_{\mathcal{N}-2,x_{2}} & \ldots & h_{\mathcal{N}-2,x_{\mathcal{N}-2}}
\end{pmatrix}
\]
is of full rank at $\left(\mathbf{0},\mathbf{0}\right)$. Then we
can make the $C_{s}$-equivariant coordinate change
\[
\tilde{x}_{1}=h_{2},\ldots,\hspace{1em}\tilde{x}_{\mathcal{N}-3}=h_{\mathcal{N}-2},\hspace{1em}\tilde{x}_{\mathcal{N}-2}=h_{1},\ldots,
\]
\[
\tilde{x}_{\mathcal{N}-1}=x_{\mathcal{N}-1},\ldots,\hspace{1em}\tilde{y}_{\mathcal{N}-3}=y_{\mathcal{N}-3}
\]
in a neighbourhood of the origin. This coordinate change brings $F$
to the form
\[
F:\begin{pmatrix}x_{1},x_{2},x_{3},\ldots,y_{\mathcal{N}-3}\end{pmatrix}\mapsto\begin{pmatrix}x_{\mathcal{N}-2},\sum_{i=1}^{\mathcal{N}-3}x_{i}y_{i}\end{pmatrix}.
\]
We see that 
\[
F^{-1}\left(0\right)=\left\{ x_{\mathcal{N}-2}=x_{1}y_{1}+x_{2}y_{2}+\ldots x_{\mathcal{N}-3}y_{\mathcal{N}-3}=0\right\} .
\]
In the case $\mathcal{N}=4$, this gives
\[
F^{-1}\left(0\right)=\left\{ x_{2}=x_{1}y_{1}=0\right\} =\left\{ x_{2}=y_{1}=0\right\} \cup\left\{ x_{2}=x_{1}=0\right\} 
\]
so the solutions of $F=0$ do not form a submanifold, but a $C_{s}$-invariant
union of two submanifolds of codimension $2$. The first, $x_{2}=y_{1}=0$,
consists entirely of points with isotropy group $C_{s}$. The second,
$x_{2}=x_{1}=0$, intersects the first submanifold and, away from
this intersection, consists of points with trivial isotropy group.
The set $F^{-1}\left(0\right)$ is a submanifold everywhere apart
from at points with $x_{1}=x_{2}=y_{1}=0$. This is our fourth example
of a confluence. This particular type of confluence has been discovered
in $HNCO$ (isocyanic acid) \cite{key-10,key-6}. In the case $\mathcal{N}=5$,
we instead have
\[
F^{-1}\left(0\right)=\left\{ x_{3}=x_{1}y_{1}+x_{2}y_{2}=0\right\} 
\]
which is yet another example of a confluence. This particular type
of confluence has not yet been discovered in a real molecule, but
ought to occur generically in penta-atomic molecules. Indeed, taking
$\mathcal{N}>4$ we get a whole family of confluences (one for each
$\mathcal{N}$) which appear to be new, and ought to occur generically.
\end{itemize}

\section{Conclusions}
Our motivation for writing this paper was the discovery of confluences of conical intersection seams in the chemical literature. It is clear that confluences owe their existence to symmetry, fundamentally deriving from rotation and inversion symmetry of $3D$ space together with permutation symmetry due to the molecule (possibly) having atoms with identical nuclear charges. We have outlined how such symmetry can be incorporated into an analysis of electronic eigenvalue crossings. In the case of triatomics, we have carried out this analysis in detail, and in the process have rediscovered all types of confluence which have appeared in the chemical literature over the past $30$ years. This same analysis yields one further type of confluence, which should occur generically in triatomics with three identical atoms, yet has not been discovered in a real molecule. It would be interesting to search for this final confluence, which arises in $D_{3h}$ symmetry when an $A_1'$ state crosses an $A_2'$ state, in a real triatomic.

\end{document}